\newlength\smallfigwidth
\documentclass[aps,prb,twocolumn,floatfix,showpacs,amsmath,amssymb]{revtex4}
\usepackage[usenames,dvipsnames]{pstricks}
\usepackage{pst-grad} % For gradients
\usepackage{pst-plot} % For axes

\usepackage{graphicx}
\def\ba{\begin{eqnarray}}
\def\ea{\end{eqnarray}}
\def\be{\begin{equation}}
\def\ee{\end{equation}}
\def\nn{\nonumber}

\begin{document}

\preprint{UFV}

\title{\bf Scattering of charge carriers in graphene induced by topological defects}

\author{J.\ M.\ Fonseca} \email{jakson.fonseca@ufv.br}
%\affiliation{Departamento de F\'isica, Universidade Federal de Vi\c cosa, Vi\c cosa, 36570-000, Minas Gerais, Brazil}
\author{W.\ A.\ Moura-Melo} \email{winder@ufv.br}
%\affiliation{Departamento de F\'isica, Universidade Federal de Vi\c cosa, Vi\c cosa, 36570-000, Minas Gerais, Brazil}
\author{A.\ R.\ Pereira} \email{apereira@ufv.br}
\affiliation{Departamento de F\'isica, Universidade Federal de Vi\c
cosa, Vi\c cosa, 36570-000, Minas Gerais, Brazil}
%\affiliation{Max Planck Institute for the Physics of Complex Systems, 01187 Dresden, Germany}

\date{June 17, 2010}

\begin{abstract}
We study the scattering of graphene quasiparticles by topological
defects, represented by holes, pentagons and heptagons. For the case
of holes, we obtain the phase shift and found that at low
concentration they appear to be irrelevant for the electron
transport, giving a negligible contribution to the resistivity.
Whenever pentagons are introduced into the lattice and the fermionic
current is constrained to move near one of them we realize that such
a current is scattered with an angle that depends only on the number
of pentagons and on the side the current taken. Such a deviation may
be determined by means of a Young-type experiment, through the
interference pattern between the two current branches scattered by a
pentagon. In the case of a heptagon such a current is also scattered
but it diverges from the defect, preventing a interference between
two beams of current for the same heptagon.
\end{abstract}

\pacs{81.05.Uw, 73.63.Fg, 04.20.-q}

\maketitle

\section{Introduction and Motivation}

\begin{figure}
\includegraphics[angle=0.0,width=6cm]{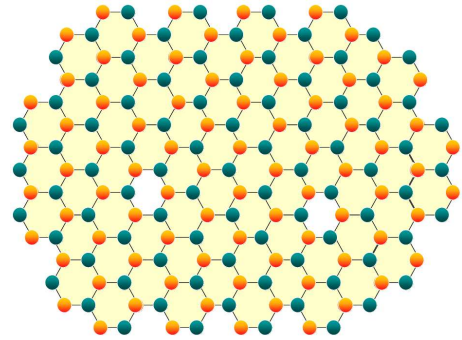}
\caption{ \label{figplano} (Color online) Graphene is a flat
monolayer of carbon atoms having a honeycomb lattice, consisting of
two interpenetrating triangular sublattices (green and red circles).
Experimental techniques provide high-quality graphene crystallites
up to $100\mu m $ in size, which is sufficient for most research
purpose, including the ones considered here. In this ``perfect''
layer the charge carries can travel thousands of interatomic
distances without scattering.}
\end{figure}

Graphene is a flat monolayer of carbon atoms tightly packed into a two-dimensional ($2D$)  honeycomb lattice, (Fig.\ \ref{figplano}) or it can be viewed as an individual atomic plane pulled out of bulk graphite \cite{Novoselov04,Novoselov05}. %[\onlinecite{Novoselov04,Novoselov05}].
It is the first example of a truly atomic two-dimensional $(2D)$
crystalline system and it is the basic building blocks for graphitic
materials such as fullerenes (graphene balled into a sphere) or
carbon nanotubes (graphene
rolled-up in cylinders) \cite{Geim07,Castro-Neto06,Geim08}. %[\onlinecite{Geim07,Castro-Neto06,Geim08}].
Additionally, it is a zero-gap semiconductor, in which the low
energy spectrum is correctly described by the
$(2+1)D$ Dirac-like equation for a massless particle \cite{Novoselov05,Katsnelson06, Wallace47} %[\onlinecite{Novoselov05,Katsnelson06,Wallace47}]

\be\label{Dirac} i\hbar\frac{\partial}{\partial t}|\Psi\rangle =
v_F\vec{\sigma}\cdot \vec{p} |\Psi\rangle,
\ee\\
where $v_{F}$ is the Fermi velocity, which plays the role of the
speed of light ($v_{F}\approx c/300$), $\vec{\sigma}=(\sigma_{x},
\sigma_{y})$ are the $2D$ Pauli matrices, $\vec{p}=-i\hbar
\vec{\nabla}$ is the linear momentum operator and $|\Psi\rangle$ is
a two-component spinor. Therefore, the quasiparticles can be viewed
as electrons that have lost their masses or as (massless) neutrinos
that acquired the electronic charge. Such a spectrum makes graphene
a material with unique electronic properties. Its description by
means of the Dirac equation is a direct consequence of graphene?s
crystal symmetry. Its honeycomb lattice is made up of two equivalent
triangular carbon sublattices $A$ and $B$, whereas its cosine-like
energy bands associated with the sublattices intersect at zero
energy $(E=0)$ near the edges of the Brillouin zone, giving rise to
a conical section spectrum
at low energies\cite{Geim07}, %[\onlinecite{Geim07}],
say, $|E|\,<\,1$ eV. In this honeycomb lattice, the two-component
spinor $|\Psi\rangle$ is referred to as pseudospin since it is an
index indicating two interpenetrating triangular sublattices $A$ e
$B$, which is similar to spin index (up and down) in quantum
electrodynamics (QED). It is common to regard the sublattice degree
of freedom as a pseudospin, with the $A$ sublattice being the ``up",
$|+ \rangle$ and $B$ sublattice being the ``down", $|- \rangle$,
i.e.,

\begin{eqnarray}\label{ABphase}
|+ \rangle=\left(
  \begin{array}{cc}
    1 \\
    0 \\
  \end{array}
\right) ;\qquad  |- \rangle=\left(
  \begin{array}{cc}
    0 \\
    1 \\
  \end{array}
\right).
\end{eqnarray}\\
Since $v_{F}\ll c$, it is a slow relativistic system or a strong
coupling version of QED since the graphene's dimensionless coupling
constant, $e^{2}/\hbar v_{F}\approx 1$ much higher than its QED
analogue, the fine structure constant $e^{2}/\hbar c\approx 1/137$.
All these properties make the graphene a very
interesting system, which provides a way to probe QED phenomena, for instance, by measuring its electronic properties. Several proposals for testing some predicted, but not yet observed phenomena in QED, including the Klein paradox \cite{Katsnelson06,Katsnelson07}, %[\onlinecite{Katsnelson06, Katsnelson07}],
vacuum polarization \cite{Shytov07} %[\onlinecite{Shytov07}]
and atomic collapse \cite{Shytov+07}, %[\onlinecite{Shytov+07}]
are some topics under investigation in graphene.

Here, we would like to study the behavior of graphene quasiparticles
in the presence of defects in the crystalline structure of the
material. We shall consider three types of defects: holes, pentagons
and heptagons. All these defects can be incorporated by removing or
inserting a few carbon atoms. Understanding how these defects modify
the transport properties of graphene is crucial to achieve future
electronic devices using carbon-made materials.

The presence of defects like pentagons (heptagons) induces positive
(negative) curvature in the material. At some extent the charge
carriers motion in the presence of pentagonal (heptagonal) defects
is identical to fermions moving in a $(2+1)D$ gravitational
space-time generated by positive (``negative'') point-like masses.
Then we may employ results from gravity to analyze some effects
concerning charge carriers in the presence of such defects. For
instance, in the presence of pentagons and heptagons we found that
the phase shifts of the carriers moving near these defects depend
only on the number of pentagons or heptagons in the graphene sheet
and are identical for charge carriers in the sublattice $A$ or $B$.
In the case of the holes we computed the phase shifts for the
scattered electrons and found that at low concentration they are
irrelevant for the electrons transport giving no appreciable
contribution  to the resistivity.

\section{Charge carriers dynamics in a non-simply connected graphene sheet}

In this section we discuss the scattering of the charge carriers in a non-simply connected graphene sheet. In the continuum model for the graphene, it is assumed that there is a hole of radius $r_0$ cut from the system center, located at the origin. Thus the motion of the quasiparticles is performed on a flat $2D$ support given by a non-simply connected manifold, which can be viewed as defect in the material.  This model allows for the investigation of scattering effects as a function of the hole radius and could shed some light on the high charge carrier mobilities, fact observed in graphene \cite{Novoselov05,Katsnelson08}. %[\onlinecite{Novoselov05}].
In addition, it is interesting to compare our results with the scattering by a short-range potential because it appears that in the case of graphene the contribution of such small defects to the resistivity is essentially smaller than for conventional nonrelativistic two-dimensional electron gas \cite{Katsnelson07,Katsnelson08}. %[\onlinecite{Katsnelson07}].

To determine the phase-shift of the scattered wave function as well as the scattering cross section one has to solve the two-dimensional Dirac equation (\ref{Dirac}) which, for the case of massless particles, can be writing in a covariant form,%\footnote{The word covariant must be used carefully because $v_F$ is not invariant, being only a parameter and the term covariant is used to refer only the form the equation is written.}
like below:

\ba\label{Dirac covariante}
i\hbar\gamma^\mu\partial_\mu\psi(x)\,=\,0,
\ea\\
where the covariant derivative is $\partial_\mu = [(1/v_F)\partial/
\partial t\,, \,\partial/\partial x\,, \,\partial/ \partial y\,]$,
the $\gamma$-matrices are $\gamma^0=\sigma^3,$ $\gamma^1=i\sigma^2$
and $\gamma^3=-i\sigma^1,$ obeying
$\gamma^\mu\gamma^\nu=\eta^{\mu\nu}-i\epsilon^{\mu\nu\alpha}\gamma_\alpha$,
$\eta^{\mu\nu}$ is the Minkowski tensor metric, ${\rm
diag}(\eta^{\mu\nu})=(+1,-1,-1)$ and $\epsilon^{\mu\nu\alpha}$ is
the 3-dimensional Levi-Civita symbol $(\epsilon^{012}\equiv+1).$
(The word covariant must be used carefully because $v_F$ is not
invariant, being only a parameter and the term covariant is used to
refer only the form the equation is written.) We may expand the
solutions of the free massless Dirac equation in plane waves, once
rotational invariance allows to separate the $\theta$ variable, so
that the diagonalized angular momentum
$\mathcal{J}=-i\hbar\frac{\partial}{\partial\theta}
+\frac\hbar2\sigma^3$, yielding partial waves with angular momentum
$(n+\frac12)\hbar$, takes the form,

\be
\psi(\vec{r},t)=e^{i(n+\frac12-\frac12\sigma^3)\theta}u_n(r)e^{-iEt/\hbar}.
\ee\\
The components of the radial spinor $u_n(r)$, given by $f_n(r)$ and
$g_n(r)$, satisfy the Bessel equation of order $n$ and $n+1$, say:

\ba\label{Bessel equations}
\frac{d^2 f_n(r)}{dr^2}+\frac1r\frac{d f_n(r)}{dr}+\bigg(k^2-\frac{n^2}{r^2}\bigg)f_n(r)\,=0\,,\\
\frac{d^2 g_n(r)}{dr^2}+\frac1r\frac{d
g_n(r)}{dr}+\bigg(k^2-\frac{(n+1)^2}{r^2}\bigg)g_n(r)\,=0\,,
\ea\\
where $n\,=\,0\,,\pm1\,,\pm2\,,\ldots$ is the angular-momentum
number and $k=\frac{E}{\hbar v_F} > 0\,$ (we are considering only
solutions with $E>0$ that describe the electronic dynamics).
Remember that the spinor index in graphene labels its two crystal
sublattices rather than directions of the actual spin. The solution
for the radial spinor outside of the hole (i.e., for $r>r_0$) is
given by:

\ba\label{general solution} u_n(r)&=&\left(\begin{array}{c}
        f_n(r) \\
        g_n(r)
      \end{array}\right)\nn\\
&=&\left(
          \begin{array}{c}
            B_{1n}\,J_n(k r) + B_{2n}\,N_n(k r)\\
            B_{3n}\,J_{n+1}(k r) + B_{4n}\,N_{n+1}(k r)\\
          \end{array}
        \right),
\ea\\
where $J_n$ and $N_n$ are the Bessel functions of first and second
kinds (Neumann function), respectively, whereas
$B_{jn}\,\,(j\,=1\,,\,2\,,\,3\,,\,4)$ are constants. These constants
must be determined by the appropriate boundary conditions specified
to completely define the problem. From the physical point of view,
the correct boundary condition is determined by the requirement of
vanishing the net energy flux into the hole, which is a region
absent of lattice degrees of freedom. Consequently, the fields must
arrange themselves in such a way that the energy flux from the
incoming modes (asymptotically behaving like $e^{-ikr}$) exactly
cancel that from the outgoing waves.

Let us recall that in the case without a hole, one has:

\be\label{solution without a hole} u_n(r)=\left(
          \begin{array}{c}
            B_{1n}\,J_n(k r)\\
            B_{3n}\,J_{n+1}(k r)\\
          \end{array}
        \right).
\ee\\
To determine the phase-shift $\delta_n$, we need the asymptotic behavior, $r\rightarrow \infty$, of the radial spinor (\ref{solution without a hole}) say\cite{Morse Feshback}: %[\onlinecite{Morse Feshback}]

\ba\label{asymptotic plane wave}
u_n^A(r)\Big|_{r\rightarrow\infty}  \longrightarrow \sqrt{\frac{2}{\pi kr}}\cos \bigg(kr-\frac{n\pi}{2}-\frac{\pi}{4}+\delta_n^A\bigg)\,,\nn \\
u_n^B(r)\Big|_{r\rightarrow\infty} \longrightarrow
\sqrt{\frac{2}{\pi kr}}\cos
\bigg(kr-\frac{(n+1)\pi}{2}-\frac{\pi}{4}+\delta_n^B\bigg)\,,
\ea\\
where the spinor indices $A$ and $B$ labels the two crystal
sublattices. Now, imposing Neumann boundary condition (NBC) on the
wave-functions, $\frac{\partial u_n}{\partial r}\Big|_{r=r_0}=0$, we
obtain:

\be\label{solution with a hole} u_n(r)=\left(
          \begin{array}{c}
            B_{1n} \big [J_n(kr)-\tan(t_n(kr_0))N_n(kr)\big]\\
            B_{3n}\big[J_{n+1}(kr)-\tan(t_{n+1}(kr_0))N_{n+1}(kr)\big] \\
          \end{array}
        \right)\,,
\ee\\
where \be
\tan[t_n(kr_0)]=\frac{J'_n(kr_0)}{N'_n(kr_0)}\,,%\qquad
%t_n(kr_0)=\arctan{\Bigg[\frac{n J_n(kr_0)-kr_0J_{n+1}(kr_0)}{n N_n(kr_0)-kr_0N_{n+1}(kr_0)}\Bigg]},
\ee\\
The terms proportional to Bessel (Neumann) functions describe the
incident (scattered) waves. Comparing their asymptotic behavior with
those for plane waves, eq. (\ref{asymptotic plane wave}), we obtain
the phase shifts. Asymptotically, the scattered wave-function
$u_n(r)$ behaves like follows:

\ba\label{asymptotic with a hole}
\hspace{-.5cm}u_n^A(r)\Big|_{r\rightarrow\infty} \hspace{-.8cm}&
\longrightarrow&\sqrt{\frac{2}{\pi kr}}
\cos\Big(kr-\frac{n\pi}{2}-\frac{\pi}{4}+t_n(kr_0)\Big)\,, \nn \\
\hspace{-.5cm}u_n^A(r)\Big|_{r\rightarrow\infty} \hspace{-.8cm}&
\longrightarrow& \hspace{-.2cm}\sqrt{\frac{2}{\pi
kr}}\cos\hspace{-.1cm}\Big(kr\hspace{-.1cm}-\hspace{-.1cm}\frac{(n\hspace{-.1cm}+\hspace{-.1cm}1)\pi}{2}\hspace{-.1cm}
-\hspace{-.1cm}\frac{\pi}{4}\hspace{-.1cm}+\hspace{-.1cm}t_{n+1}(kr_0)\hspace{-.1cm}\Big)\,,
\ea\\
giving us the phase-shift $\delta_n$ of the n-th partial wave that
completely determines the fermionic scattering:

\be \delta_n^A = t_n(kr_0)\,, \ee
\be \delta_n^B = t_{n+1}(kr_0)\,.
\ee\\

Now, if we consider a small concentration of point-like defects with concentration $n_{\rm def}$ angle-depedent scattering cross section, $\sigma(\theta),$ their contribution to resistivity, $\rho$, reads \cite{Katsnelson07}:%[\onlinecite{Katsnelson07}]:

\ba\label{resistivity}
\rho&=&\frac{2}{e^2v_F^2N(E_F)\tau(k_F)}\,,\nn\\
\frac{1}{\tau(k_F)}&=&n_{\rm def}
v_F\int_0^{2\pi}d\theta\frac{d\sigma(\theta)}{d\theta}(1-\cos\theta)\,,
\ea\\
where $N(E_F)=2k_F/\pi\hbar v_F$ is the density of states at the
Fermi level, taking into account the four-fold degeneracy, the spin
and valley of the graphene, and $\tau$ is the mean-free path. For
electrons of the sublattice $A$ we have:

\be\label{diferencial cross section}
\frac{d\sigma(\theta)}{d\theta}=\frac{2}{\pi
k}\Bigg|\sum_{n=-\infty}^{n=\infty}t_n(kr_0)e^{i\theta}\Bigg|^2\,.
\ee\\
The Dirac eq. (\ref{Dirac covariante}) or (\ref{Bessel equations})
has an important symmetry under the interchanging
$f\longleftrightarrow g\,,\, n\longleftrightarrow -n-1$ which
implies that $t_n=t_{-n-1}$. Thus, eq. (\ref{diferencial cross
section}) can be written in the form:

\be\label{diferencial cross section 2}
\frac{d\sigma(\theta)}{d\theta}=\frac{2}{\pi
k}\Bigg|\sum_{n=0}^{n=\infty}t_n(kr_0)\cos[(n+1/2)\theta]\Bigg|^2\,.
\ee\\
For small energies, $kr_0 \ll1$, which is typical for graphene, one
has

\be\label{phase shifts limit} \delta_n^A = t_n(kr_0) \simeq
\frac{(2n+1)[(2n+3)n-(kr_0)^2]}{[(2n+1)!!]^2(2n+3)(n+1)}(kr_0)^{2n+1}
\ee\\
and thus the s-scattering $(n=0)$ dominates. With eqs.
(\ref{diferencial cross section 2}-\ref{phase shifts limit}) the
impurities contribution to the resistivity may be estimated as

\be\label{resistivity with a hole}
\rho%=n_{\rm def}\frac{h}{2\pi^2e^2kk_F}(kr_0)^6
\simeq n_{\rm def}\frac{h}{e^2kk_F}(kr_0)^6\,.
\ee\\
This means that the scattering induced by small holes (with radius around a few angstroms; some lattice spacings) at low concentration are irrelevant for the electronic transport in graphene, giving a negligible contribution to the resistivity. For the case of a potential $V(r)=V_0$ at $r<R_0$ and $V(r)=0$ at $r>R_0,$ the estimation for this type of impurity contribution to the resistivity is\cite{Katsnelson07} %[\onlinecite{Katsnelson07}]
$\rho\simeq(h/4e^2)n_{\rm def}R_0^2$, giving a negligible
contribution to the resistivity when the radius of the potential
$R_0$ is of the order of interatomic distances and at a low
concentration as above.

For an intuitive understanding of the result (\ref{resistivity with a hole}) let us recall that light does not experience obstacles with sizes much smaller than its wavelength, whereas massless Dirac electrons have the same dispersion relation behaving as light in some aspects (duality wave-particle). This same interpretation can explain the results for a short range potential \cite{Katsnelson08}. %[\onlinecite{Katsnelson08}].
Ours results for the resistivity are in agreement with those
presented in Ref.[\onlinecite{Katsnelson08}], where the authors show
that intrinsic corrugations of a graphene sheet create a long-range
scattering potential and lead to significant resistivity that could
explain the existing experimental data about the graphene
resistivity (to more details see Ref.[\onlinecite{Katsnelson08}]).
For more details about scattering of the charge carriers by defects
in graphene see, for instance
[\onlinecite{Katsnelson08,Rutter,Katsnelson09,Katsnelson07 bilayer}]
and references therein.

\section{Charge-carriers dynamics in the presence of pentagonal and heptagonal defects}

Here, pentagons and heptagons defects are considered and their effects in the charge carriers dynamics is investigated. Substitution of an hexagon by other type of polygon with $n\,=\,6-n_d$ sides, where $n_d$ is an integer smaller than $6$, in the lattice without affecting the threefold coordination of the carbon atoms leades to the wrapping of the graphene sheet. These defects can be seen as disclinations of the lattice locally acquiring  a finite curvature. The accumulations of various defects may lead to closed shapes as fullerenes. Rings with $n<6$ sides $(n_d>0)$ give rise to positively curved structures whereas polygons with $n>6$ sides $(n_d<0)$ yields negative curvature. This induced curvature holds only near the defect, whereas the graphene sheet is flat away from the defect itself, as the conical surface is out from the apex \cite{Furtado08,Vozmediano07,Sitenko07}. %[\onlinecite{Furtado08,Vozmediano07,Sitenko07}].

\begin{center}
\begin{figure}[h!] \label{foto1-UFV}
\includegraphics[angle=0,width=4cm]{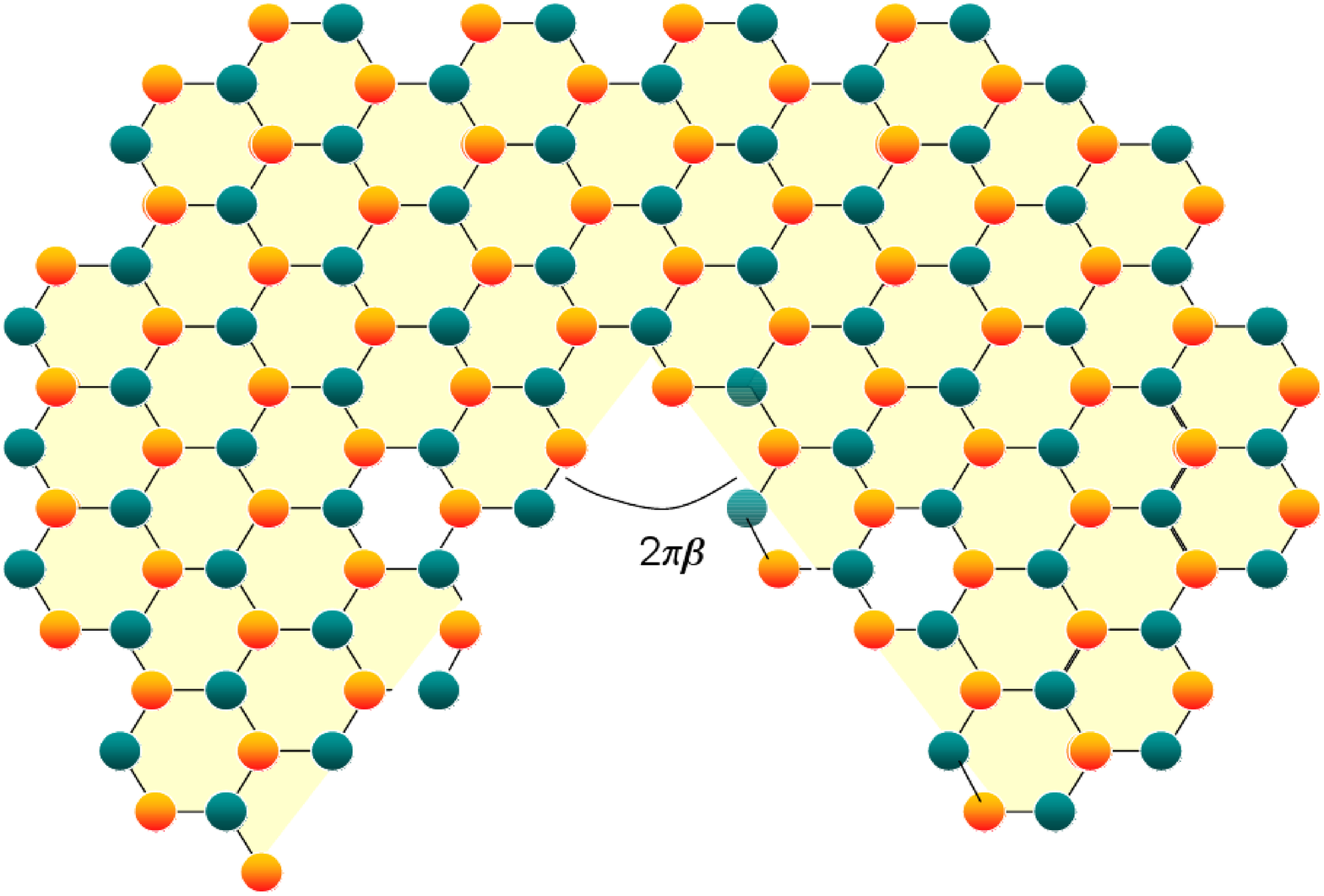} \hskip .5cm
\includegraphics[angle=0,width=4cm]{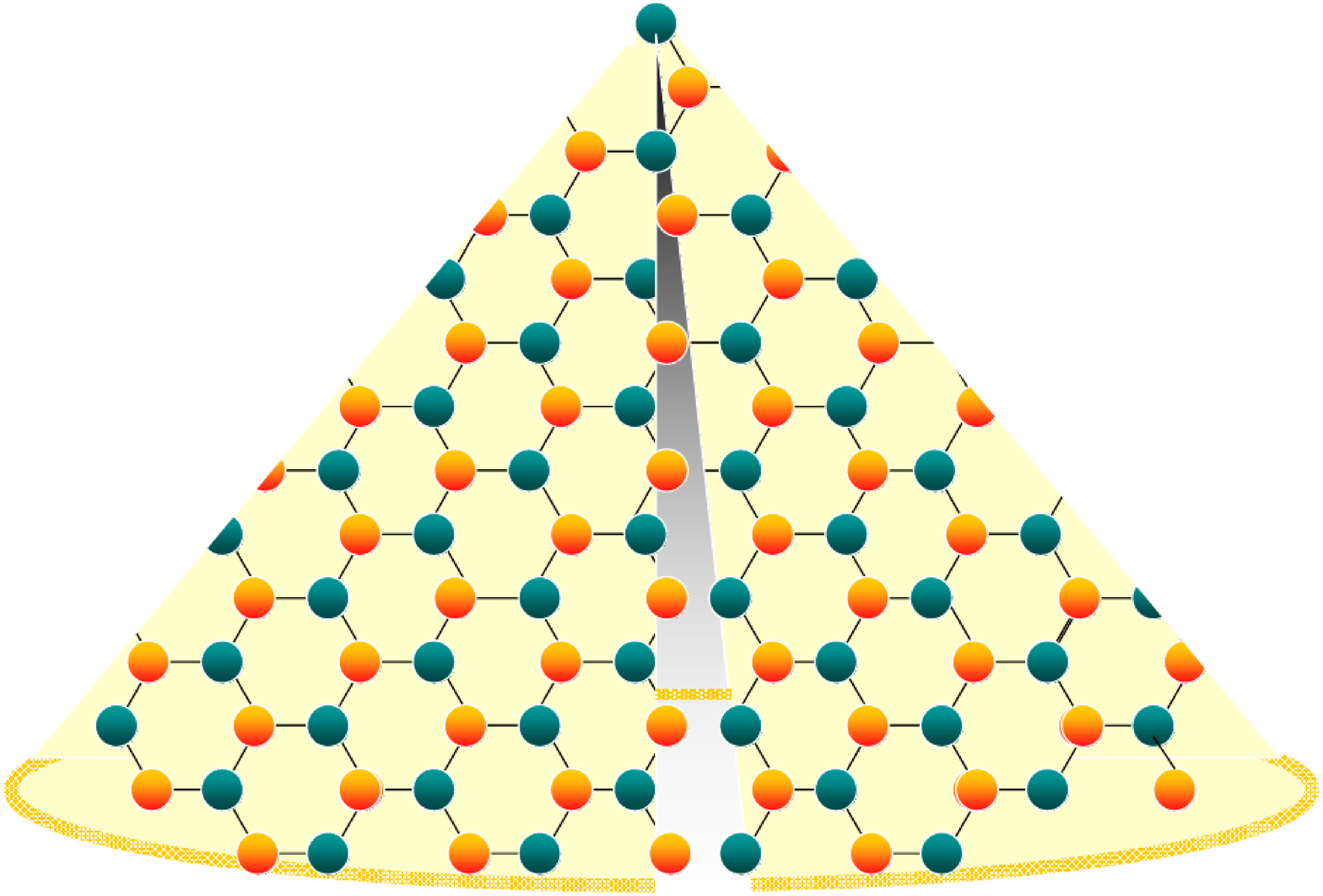}
\caption{ \label{figcone}  (Color online) Removing a wedge from the
graphene (left) and identifying the edges (right), a cone results.
The motion of the charge carriers in an ideal conical graphene is
equivalent to that of a massless Dirac particle in a gravitational
field of a static particle of mass $M$ in a $(2+1)D$ space-time.}
\end{figure}
\end{center}

Exploring the $2D$ character and flexibility of this material, our
idea is to propose a system which one or more sectors are excised
from a graphene and the remainder is joined seamlessly, (Fig.\
\ref{figcone}). In fact, the missed link of each carbon atom resting
at the two edges of the remaining
graphene sheet can be, in principle, covalently bounded. The nucleation and growth of curved carbon structures remain to be well-understood. It is claimed that the occurrence of pentagons, which yields $60^{\circ}$ disclination defects in hexagonal graphitic network, is a key element in this scenario. Particularly, considering the symmetry of a graphitic sheet and the Euler theorem, it can be shown that only five types of cones (incorporating one to five pentagons) can be made of a continuous graphene sheet \cite{Ge94,Krishnan97}. %[\onlinecite{Ge94,Krishnan97}].
In the case of a cone with $n_d>0$, the value $n_d$,
$(n_d\,=\,1\,,\ldots\,,\,5)$ is related to the conical angle
$\gamma$:

\be\label{angulosuperficie} \sin\frac\gamma2=1-\frac{n_d}{6}\,.
\ee\\
The deficit-angle induced by the conical singularity is given by $2\pi(1-n_d/6)$. The pentagonal defect can be presented as a pseudo-magnetic vortex at the apex of a graphitic cone, being the flux of the vortex related to the deficit angle of the cone (see Ref. \cite{Sitenko07}). %[\onlinecite{Sitenko07}].
The possible five graphitic cones mentioned earlier are then given by $\gamma=19.2^{\circ}, 38.9^{\circ}, 60^{\circ}, 84.6^{\circ}, 112.9^{\circ}$ \cite{Ge94,Krishnan97}. %[\onlinecite{Ge94,Krishnan97}].
Cones with a heptagon have negative curvature and are obtained by a
insertion of a angular sector in the carbon sheet. Then if $n_d<0$,
$-n_d$ counts the number of such sectors inserted into the graphene
sheet. Our aim is, therefore, to see the influences that such a
special graphene structure could induce on quasiparticle
wavefunctions (spinors); surely, these influences may create new
perspectives in the electronic transport properties, which are
determined by the quasiparticles constrained to move on the conical
surface. Once only the honeycomb lattice ensures the graphene
peculiar dispersion relation $E=v_Fp$, lattice distortions
deviations should be minimized as much as possible wherever building
the cones, say, the number of inserted pentagons must be kept a
minimum.

%\begin{figure}
%\includegraphics[angle=0.0,width=6cm]{figremove.eps}
%\caption{ \label{figremove} (Color online) Detaching a fraction of
%a graphene sheet to generate a deficit angle of $2\pi \beta$.}
%\end{figure}

%\begin{figure}
%\includegraphics[angle=0.0,width=6cm]{figcone.eps}
%\caption{ \label{figcone}  (Color online) Removing a wedge
%from the graphene (see Fig. \ \ref{figremove}) and identifying the edges, a cone results. The motion of the charge carriers in
%an ideal conical graphene is equivalent to that of a massless
%Dirac particle in a gravitational field of a static particle of
%mass $M$ in a $(2+1)D$ space-time.}
%\end{figure}

To study the scattering of the carriers in graphene by topological defects we employ the analogy between defects in condensed matter physics and in $(2+1)$-dimensional gravity \cite{Volovik03} %[\onlinecite{Volovik03}]
as far as possible. For example, the dynamics of charge carriers  in an ideal conical graphene is equivalent to that of a massless Dirac particle in a gravitational field of a static point-like mass in a $(2+1)D$ space-time\cite{Furtado08,Brown88,Souza89}. %[\onlinecite{Furtado08,Brown88,Souza89}].
Specifically, we shall consider one of the simplest curved manifold,
which is associated to the Schwarzschild solution in $(2+1)$
dimensions: a space-time locally flat with global nontrivial
properties described below.

To proceed further, we summarize some aspects of general relativity in three space-time dimensions. It fundamentally differs from its 4-dimensional counterpart. Indeed, it exhibits some unusual features, which can be deduced from the properties of the Einstein field equations and the Riemann curvature tensor $R^{\mu}_{\;\;\upsilon \varepsilon \kappa}$ \cite{Brown88,Staruszkiewikz63}. %[\onlinecite{Brown88,Staruszkiewikz63}]
In regions free of matter (where the momentum-energy tensor $T^{\mu
\nu}$ vanishes), the space-time is locally flat when the
cosmological constant is zero (the Einstein tensor $G^{\mu\nu}$
vanishes). However, this does not mean that a massive source has no
gravitational effects: a light beam passing by a massive, point-like
mass will be deflected
\cite{Staruszkiewikz63,Deser84,Giddings84,Vilenkin81}
%[\onlinecite{Staruszkiewikz63,Deser84,Giddings84,Vilenkin81}]
and parallel transport in a closed circuit around it will in generally gives nontrivial results \cite{Burges85,Bezerra87}. %[\onlinecite{Burges85,Bezerra87}].
Indeed, while the local curvature vanishes outside the sources,
there are nontrivial global effects. For instance, for the special
case of a point-like mass, $M$, sitting at rest in the origin, the
line element is given by

\be\label{metricflat} ds^{2}=
dt^{2}-d\rho^{2}-\rho^{2}\alpha^{2}d\theta^{2},
\ee\\
with $0\leq \theta < 2\pi $ and $\alpha=1-4GM$ ($G$ is the Newton's constant, with dimensions of $[length]^1$ instead of $[length]^2$, in natural units $\hbar=c=1$). Note that although the situation looks trivial, the coordinate $\theta$ ranges from $0$ to $2\pi \alpha$, indicating an angular deficit in space. Then, the spatial part of the metric is that of a plane with a wedge removed and edges identified; the unique $2D$ spatial geometry satisfying this description is the cone \cite{Staruszkiewikz63}. %[\onlinecite{Staruszkiewikz63}].

Alternatively, one may use embedded coordinates $r$ and $\theta$ in the three-dimensional Euclidian space which extend over the complete range, $0\leq r \leq\infty\,,\, 0\leq\theta\leq 2\pi$, and describe a cone with the constraint $z=\sqrt{(\alpha^{-2}-1)(x^2+y^2)}$, being the line element given by \cite{Souza89}: %[\onlinecite{Souza89}]:

\be\label{metricimbedded} ds^2=dt^2-\alpha^{-2}dr^2-r^2d\theta^2.
\ee\\
The attributes of the source are coded in the global properties of
the locally flat variables. All the information lies in the
non-trivial boundary conditions, which is important for the quantum
scattering of graphene charge carriers by defects, like pentagons
and heptagons.

In the case of a topological defect in graphene, it is useful to
change the gravitational term $4GM$ by the symbol $\beta$, so that
$2\pi \beta$ (for $0<\beta<1$) gives the deficit of angle measuring
the magnitude of the removed sector whereas $-2\pi\beta$ (for
$-\infty<\beta<0$) accounts for the angle in excess associated to
the insertion of a sector. The parameter $\beta$ takes only discrete
values because of the lattice symmetry of the graphene as discussed
after eq. (\ref{angulosuperficie}).

%%%%%%%%%%%%%%%%%%%%%%%% FIGURA $$$$$$$$$$$$$$$$$$$$$$$$$$$$$$$$$

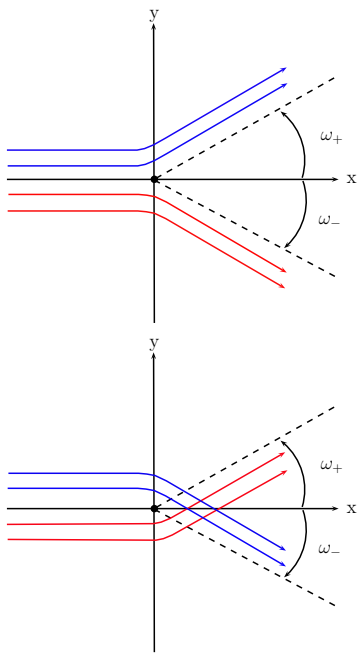
\begin{figure}
\begin{center}
\scalebox{0.5} % Change this value to rescale the drawing.
{
\begin{pspicture}(0,-8.569219)(10.241875,8.589219)
\definecolor{color121}{rgb}{0.058823529411764705,0.027450980392156862,0.9647058823529412}
\definecolor{color124}{rgb}{0.09411764705882353,0.0196078431372549,0.9607843137254902}
\definecolor{color127}{rgb}{0.996078431372549,0.023529411764705882,0.023529411764705882}
\definecolor{color130}{rgb}{0.9803921568627451,0.054901960784313725,0.054901960784313725}
\definecolor{color154}{rgb}{0.9686274509803922,0.0196078431372549,0.09803921568627451}
\definecolor{color157}{rgb}{0.9725490196078431,0.027450980392156862,0.09411764705882353}
\definecolor{color160}{rgb}{0.047058823529411764,0.011764705882352941,0.984313725490196}
\definecolor{color163}{rgb}{0.050980392156862744,0.01568627450980392,0.9411764705882353}
\psline[linewidth=0.04cm,arrowsize=0.05291667cm
2.0,arrowlength=1.4,arrowinset=0.4]{->}(0.0,4.0307813)(8.82,4.0307813)
\psline[linewidth=0.04cm,fillcolor=black,linestyle=dashed,dash=0.16cm
0.16cm,dotsize=0.07055555cm 2.0]{**-}(3.84,3.9907813)(8.7,6.730781)
\psline[linewidth=0.04cm,linestyle=dashed,dash=0.16cm
0.16cm](3.92,4.0107813)(8.72,1.4507812)
\psline[linewidth=0.04cm,arrowsize=0.05291667cm
2.0,arrowlength=1.4,arrowinset=0.4]{<-}(3.9,8.190782)(3.92,0.21078125)
\psline[linewidth=0.04,linecolor=color121,arrowsize=0.05291667cm
2.0,arrowlength=1.4,arrowinset=0.4]{->}(0.02,4.3907814)(3.48,4.3907814)(3.68,4.4307814)(3.92,4.5307813)(4.24,4.710781)(7.46,6.590781)
\psline[linewidth=0.04,linecolor=color124,arrowsize=0.05291667cm
2.0,arrowlength=1.4,arrowinset=0.4]{->}(0.0,4.8107815)(3.46,4.8107815)(3.66,4.8507814)(3.9,4.9507813)(4.22,5.130781)(7.44,7.0107813)
\psline[linewidth=0.04,linecolor=color127,arrowsize=0.05291667cm
2.0,arrowlength=1.4,arrowinset=0.4]{->}(0.04,3.6307812)(3.5,3.6307812)(3.72,3.610589)(3.94,3.5707812)(4.24,3.4288583)(7.44,1.5507812)
\psline[linewidth=0.04,linecolor=color130,arrowsize=0.05291667cm
2.0,arrowlength=1.4,arrowinset=0.4]{->}(0.02,3.1907814)(3.48,3.1907814)(3.7,3.170589)(3.92,3.1307812)(4.22,2.9888582)(7.4,1.1307813)
\rput{-127.32463}(6.481249,12.287562){\psarc[linewidth=0.04,arrowsize=0.05291667cm
2.0,arrowlength=1.4,arrowinset=0.4]{->}(6.2820997,4.5395117){1.6364056}{110.83665}{180.0}}
\rput{-161.10881}(11.183567,8.779596){\psarc[linewidth=0.04,arrowsize=0.05291667cm
2.0,arrowlength=1.4,arrowinset=0.4]{<-}(6.3220997,3.4595118){1.6364056}{110.83665}{180.0}}
\rput(3.925625,8.3957815){\Large y} \rput(9.166562,4.0357814){\Large
x} \rput(8.651406,5.1357813){\Large $\omega_{+}$}
\rput(8.591406,2.8957813){\Large $\omega_{-}$}
\psdots[dotsize=0.2](3.92,4.0307813)
\psline[linewidth=0.04cm,arrowsize=0.05291667cm
2.0,arrowlength=1.4,arrowinset=0.4]{->}(0.0,-4.729219)(8.82,-4.729219)
\psline[linewidth=0.04cm,fillcolor=black,linestyle=dashed,dash=0.16cm
0.16cm,dotsize=0.07055555cm
2.0]{**-}(3.84,-4.769219)(8.7,-2.0292187)
\psline[linewidth=0.04cm,linestyle=dashed,dash=0.16cm
0.16cm](3.92,-4.749219)(8.72,-7.309219)
\psline[linewidth=0.04cm,arrowsize=0.05291667cm
2.0,arrowlength=1.4,arrowinset=0.4]{<-}(3.9,-0.56921875)(3.92,-8.549219)
\psline[linewidth=0.04,linecolor=color154,arrowsize=0.05291667cm
2.0,arrowlength=1.4,arrowinset=0.4]{->}(0.0,-5.1492186)(3.9,-5.1292186)(4.1,-5.0892186)(4.34,-4.9892187)(4.66,-4.809219)(7.4,-3.2292187)
\psline[linewidth=0.04,linecolor=color157,arrowsize=0.05291667cm
2.0,arrowlength=1.4,arrowinset=0.4]{->}(0.02,-5.5492187)(3.96,-5.5692186)(4.16,-5.5292187)(4.4,-5.429219)(4.72,-5.249219)(7.46,-3.7092187)
\psline[linewidth=0.04,linecolor=color160,arrowsize=0.05291667cm
2.0,arrowlength=1.4,arrowinset=0.4]{->}(0.02,-4.1892185)(3.48,-4.1892185)(3.7,-4.209411)(3.92,-4.249219)(4.22,-4.391142)(7.42,-6.269219)
\psline[linewidth=0.04,linecolor=color163,arrowsize=0.05291667cm
2.0,arrowlength=1.4,arrowinset=0.4]{->}(0.04,-3.7892187)(3.5,-3.7892187)(3.72,-3.809411)(3.94,-3.8492188)(4.24,-3.9911418)(7.42,-5.849219)
\rput{-127.32463}(13.447313,-1.7838916){\psarc[linewidth=0.04,arrowsize=0.05291667cm
2.0,arrowlength=1.4,arrowinset=0.4]{->}(6.2820997,-4.220488){1.6364056}{110.83665}{180.0}}
\rput{-161.10881}(14.01981,-8.268547){\psarc[linewidth=0.04,arrowsize=0.05291667cm
2.0,arrowlength=1.4,arrowinset=0.4]{<-}(6.3220997,-5.300488){1.6364056}{110.83665}{180.0}}
\rput(3.925625,-0.36421874){\Large y}
\rput(9.166562,-4.724219){\Large x}
\rput(8.651406,-3.6242187){\Large $\omega_{+}$}
\rput(8.591406,-5.8642187){\Large $\omega_{-}$}
\psdots[dotsize=0.2](3.92,-4.729219)
\end{pspicture}
}
\end{center}
%%%%%%%%%%%%%%%%%%%%%%%%  FIM DA FIGURA $$$$$$$$$$$$$$$$$$$$$$$$$$$$$$$$$
%\includegraphics[angle=0.0,width=8cm]{figremove.eps}
\caption{ \label{classical trajectories} (Color online) Classical
trajectories (bue and red) of the scattered particles deflecteds by
heptagons (above) and pentagons (below). The trajectories are the
asymptotic motion on cone projected onto $x-y$ plane of the
embedding three dimensional space, see the text.}
\end{figure}

Before analyzing the quantum mechanical scattering by conical defects (pentagons and heptagons) let us make a digression concerning scattering of the charge carriers in graphene as classical relativistic particles. The classical equation of motion, determined by relativistic geodesic equation for the particles in a cone reads $\ddot{x} + \Gamma^\mu_{\alpha\beta} \dot{x}^\alpha\dot{x}^\beta=0,$ where the overdot indicates differentiation with respect to any convenient affine variable $\tau$ that parametrizes the path $x^\mu(\tau)$ \cite{Souza89}. %[\onlinecite{Souza89}]
The angle of scattering $\pm\omega$ for the motion of the  particles in a cone  can be obtained by integration of the classical equations of motion and is given by \cite{Souza89}. %[\onlinecite{Souza89}]:

\be\label{angle scattering classical} \pm \omega=\pm
\pi(\alpha^{-1}-1)=\pm \pi\frac{\beta}{1-\beta},
\ee\\
where $\pm$ refers to the side the charge carriers (current)
trajectory pass around the defect, see (Fig.\ \ref{classical
trajectories}). Note that the result above is valid for all values
of $\beta$ despite the sector was removed or inserted. The
scattering angle above, presented in the embedded coordinate system
(\ref{metricimbedded}) measures the deflection of the asymptotic
motion on cone projected onto $x-y$ plane of the embedding three
dimensional space. The result above suggests that a pentagon or
heptagon may be used for deviating the planar current in graphene.

To obtain the correct current deviations in graphene we have to solve the Dirac equation (\ref{Dirac}) defined in a cone, say \cite{Souza89}: %[\onlinecite{Souza89}]:

\be i\hbar\gamma^\mu E_a\,^\mu D_\mu\psi=0\,,
\ee\\
where $D_\mu=\partial_\mu+\frac12\omega_{\mu;ab}\sigma^{ab}$, $\sigma^{ab}=\frac14[\gamma^a,\gamma^b]$, and $E_a\,^\mu$ is the {\it dreibein} in coordinates $(t,r,\theta)$. The spin connection $\omega_{\mu;ab}=-\omega_{\mu;ba}$ may be written in three dimensions as $\omega_{\mu;ab}=\epsilon_{abc}\omega_\mu\,^c$ with $\epsilon_{abc}$ the Levi-Civita symbol as before \cite{Souza89}. %[\onlinecite{Souza89}].
The rotational invariance of the problem enables us to choose
positive energy solutions that are simultaneously angular momentum
eigenfunctions, with eigenvalue $(n+\frac12)\hbar$:

\be
u_n(r)e^{-iEt/\hbar}=e^{i(n+\frac12-\frac12\sigma^3)\theta}\left(
                                                              \begin{array}{c}
                                                                u_n^A(r) \\
                                                                u_n^B(r) \\
                                                              \end{array}
                                                            \right)e^{-iEt/\hbar}\,,
\ee\\
where $n=0,\pm 1, \pm 2,\ldots\, .$ The solutions for $E>0$ are \cite{Souza89}: %[\onlinecite{Souza89}]:

\be u_n^A(r)=(\epsilon_n)^n J_\nu(\kappa r)\,, \ee \be
u_n^B(r)=(\epsilon_n)^{n+1} J_\nu(\kappa r).
\ee\\
Here, $J_\nu$ is the Bessel function of order $\nu\equiv
\epsilon_n/\alpha(n+(1\mp\alpha)/2)$, $n=0,\,\pm1,\,\pm2\,,\ldots,$
$\kappa=E/\hbar v_F \alpha\,,E> 0\,,\epsilon_n=\pm1$ and the same
sign has to be chosen for the upper and lower components of the
spinor $u_n(r)$. For $0<\alpha \leq 1$ or $0< \beta \leq 1$
(remember that $\alpha=1-\beta$) we must choose $\epsilon_n = {\rm
sign}(n+(1-\alpha)/2)={\rm sign}\,n\,({\rm sign}\,0\equiv\,1)$ to
have both components regular at the origin.
%{\bf And for $-\infty\leq \alpha<0,$ we must choose }
The asymptotic form of the Bessel functions determines  the phase shifts (they are identical for the upper and lower components)and are given by \cite{Souza89}: %[\onlinecite{Souza89}]:

\ba\label{phase shift cone}
\delta_n&=&-\epsilon_n\frac{\pi}{2\alpha}((1-\alpha)n
+(1-\alpha)/2)\nn\\
&=&-\frac{\epsilon_n}{2}\pi\frac{\beta}{1-\beta}\Big(n-\frac12\Big)\,,
\ea \be \epsilon_n = {\rm sign}(n+(1-\alpha)/2)={\rm
sign}(n+\beta/2)\,.
\ee\\
The phase-shifts depend only on the number of sectors removed or
inserted in the graphene sheet, accounted by $\alpha=1-\beta$. If
$-\infty<\beta<0$, we need to be careful because $\epsilon_n = \pm
1$ depending upon the value of $(n+\beta/2)$ (but the phase-shifts
remains as above) and the phase-shifts depends only on the number of
sectors (heptagons) inserted in the flat graphene sheet. In the
presence of heptagons the carriers dynamics is identical to the that
movement of the electrons in the gravitational field of a negative
mass (although not possible in gravitation, this is feasible in the
present context).

Note that the phase shift (\ref{phase shift cone}) measures the
deflection of the asymptotic motion on the cone projected onto $x-y$
plane being qualitatively identical to the classical case discussed
before. When there is a pentagon into the lattice and the fermionic
current is constrained to pass around and sufficiently close to it
such a current is scattered by the defect with an angle which
depends only on the number of sectors removed in the graphene and on
the side current passed (see Fig.\ \ref{classical trajectories}).
After passing by the pentagon the scattered current trajectories
cross and yields an interference pattern. In the case of a heptagon,
such a current is scattered but the trajectories diverge each other.

The presence of pentagons or heptagons in the sheet of graphene may
manifest as fluctuations in the concentration of charge carriers,
modifying several physical properties. For instance, in a planar
graphene, it is known that the hopping of electrons between
sublattices produces an effective magnetic field which is
proportional, in magnitude and direction, to the momentum measured
from the Brillouin-zone corners. This effective field, which acts on
the pseudospin, may suffer important changes in the conical graphene
because a hopping of a quasiparticle, which was previously, say, in
the $A$ sublattice, will make it to become out of phase with all
quasiparticles occupying the $B$ sublattice.

%Thus, measuring this mismatch between the spinor components or its effects in the transport properties of the charge carriers, such a system would provide an interesting place to probe some predictions of the Einstein theory of gravitation in two spatial dimensions.

\section{Conclusions}

We have studied the scattering of graphene quasiparticles by
topological defects like holes, pentagons and heptagons. We obtain
the phase shift of the wave-function in all cases. For the case of
holes, the main contribution concerns the $s$ scattering and even in
this case they do not change the resistivity of the sample, at least
at low concentrations (like occurs to short range potential
impurities).
%and we found that they giving a negligible contribution to the resistivity at low concentration being not important for the electronic transport in graphene. This results are similar to the case of a short range potential. When pentagons and heptagons are introduced they deform a graphene sheet in a conical surface. Such a system makes possible the experimental study of relativistic massless quasiparticles with charge $e$ on a two-dimensional surface of a cone, or equivalently, in the ``gravitational field'' (deficit angle) of a ``point-like particle'' in $(2+1)$ dimensions. This surface is locally flat being described by the Minkowski metric, like the plane but is globally nontrivial, with
%all the information lying in the non-trivial boundary conditions, which is important for the quantum mechanical scattering of graphene charge carriers by defects, like pentagons and heptagons.

We realize that when the fermionic current is constrained to move
near and around of pentagons and heptagons introduced in the
lattice, it is scattered with an angle that depends only on the
number of defects and on which side the current taken. Such a
deviation may be determined by means of a Young-type experiment,
through the interference pattern between the two currents scattered
by the pentagon. In the case of a heptagon such a current is also
scattered but it diverges from the defect. Perhaps these effects
could be used to build channels of currents in future electronic
devices using graphene. In addition, graphene would provide an
appealing way to experimentally explore general relativity in two
spatial dimensions since such effects are predicted by this theory
\cite{Furtado08,Souza89,Burges85,Bezerra87}.
%A deficit angle incorporated in the material, induced by a defect like a pentagon, may have important influences on the electronic transport properties in ``conical graphene'' changing the local density of states and Fermi velocity \cite{Vozmediano07}, %[\onlinecite{Vozmediano07}],
%or other predicted by Aharonov-Bohm-like effect. In principle, this effect could be detected by interference experiments in structured materials and consequently these small systems could provide important probes for general relativity in dimensions smaller than four, by means of analogue models of condensed matter.

Conical spaces generated by topological defects like vortices have been recently considered and the quantum mechanical scattering of a nonrelativistic particlehave been considered in the work of Ref. \cite{sitenko}%[\onlinecite{sitenko}].
with results qualitatively similar to those obtained here.

\section{Acknowledgments}

The authors are grateful to C. furtado for having drawn their
attention to important references and for discussion. They also
thank CNPq, FAPEMIG and CAPES (Brazilian agencies) for financial
support.

\end{document}